\documentclass[twocolumn,aps,showpacs,amsmath,amssymb,prb]{revtex4}

\usepackage{graphicx}
\usepackage{mathrsfs}
\usepackage{amsmath}
\usepackage{psfrag}
\usepackage{multirow}
\usepackage{color}

\begin{document}
\preprint{APS/123-QED}

\title{Thermopower measurements on pentacene transistors\\}

\author{Adrian von M\"uhlenen}
 \affiliation{Laboratoire d'Opto\'electronique des Mat\'eriaux Mol\'eculaires, STI - IMX - LOMM, Station 3,\\ \'Ecole Polytechnique F\'ed\'erale de Lausanne, CH - 1015 Lausanne, Switzerland} 
\author{Nicolas Errien}
 \affiliation{Laboratoire d'Opto\'electronique des Mat\'eriaux Mol\'eculaires, STI - IMX - LOMM, Station 3,\\ \'Ecole Polytechnique F\'ed\'erale de Lausanne, CH - 1015 Lausanne, Switzerland}
\author{Michel Schaer}
 \affiliation{Laboratoire d'Opto\'electronique des Mat\'eriaux Mol\'eculaires, STI - IMX - LOMM, Station 3,\\ \'Ecole Polytechnique F\'ed\'erale de Lausanne, CH - 1015 Lausanne, Switzerland}
 \author{Marie-No\"elle Bussac}%
 \altaffiliation[Permanent address: ]{Centre de Physique Th\'eorique, \'Ecole polytechnique - 91128 Palaiseau Cedex, France}
 \affiliation{Laboratoire d'Opto\'electronique des Mat\'eriaux Mol\'eculaires, STI - IMX - LOMM, Station 3,\\ \'Ecole Polytechnique F\'ed\'erale de Lausanne, CH - 1015 Lausanne, Switzerland}
\author{Libero Zuppiroli}	
 \email[Author to whom correspondence should be addressed: ]{libero.zuppiroli@epfl.ch}
 \affiliation{Laboratoire d'Opto\'electronique des Mat\'eriaux Mol\'eculaires, STI - IMX - LOMM, Station 3,\\ \'Ecole Polytechnique F\'ed\'erale de Lausanne, CH - 1015 Lausanne, Switzerland}
\date{\today}

\begin{abstract}
 We present the first thermoelectric measurements on pentacene field effect transistors. We report high values of the Seebeck coefficient at room temperature between $240$ and $500 \, \mathrm{\mu V/K}$ depending on the dielectric surface treatment. These values are independent of the thickness of the channel and of the applied gate voltage. Two contributions to the Seebeck coefficient are clearly identified: the expected contribution that is dependent on the position of the transport level and reflects the activated character of carrier generation, and an unexpected intrinsic contribution of $265 \pm 40 \, \mathrm{\mu V/K}$ that is independent of the temperature and the treatment of the oxide surface. This value corresponds to an unusually large lattice vibrational entropy of $3 k_\text{B}$ per carrier. We demonstrate that this intrinsic vibrational entropy arises from lattice hardening induced by the presence of the charge-carrier. Our results provide direct evidence of the importance of electronic polarization effects on charge transport in organic molecular semiconductors.
\end{abstract}

\keywords{Organic semiconductor, charge transfer, thermoelectric power, Seebeck effect, pentacene, organic field effect transistor (OFET)}
\pacs{72.80 Le - 72.20 Pa - 33.15 Kr}
\maketitle

\section{Introduction}
 Due to the narrow-band nature of their electronic states, charge carriers in organic molecular semiconductors are prone to perturbations by electronic and lattice excitations. \cite{silinsh1994} Despite decades of research devoted to understanding the implications of these interactions to charge transport, a clear consensus is yet to emerge. The current wide interest in these materials for potential optoelectronic applications (see the review articles in Ref.~\onlinecite{faupel04}) exacerbates the need for gaining insight into this problem. Recent papers have emphasized the role of transfer-integral fluctuations \cite{Orlandi06} or polarization fluctuations \cite{picon04,houili06} in the transport process, particularly in the context of experimental observations of the decline of the carrier mobility with increasing gate-dielectric permittivity seen in both single-crystal and thin-film organic transistor devices. \cite{morpurgo04,veres03} \\
 
 Thermoelectric measurements, because they provide direct access to the heat transported by the carrier, can unravel the origin of charge transport. Surprisingly, to our knowledge, Seebeck coefficients have never been measured in the crystalline organic semiconductors such as pentacene, rubrene and related compounds of interest here.\\
 
 In this article, we present what, to our knowledge, are the first results of thermopower measurements on pentacene thin transistors. We find that a temperature-independent, intrinsic contribution to the Seebeck coefficient of $265 \pm 40 \, \mathrm{\mu V/K}$ is present in all devices regardless of the gate-insulator used. This value corresponds to a very large lattice vibrational entropy of about $3 k_\text{B}$ per carrier. We shall show that this result arises from the creation of an intermolecular phonon cloud associated with local changes in the Van-der-Waals interactions between molecules that accompanies the electronic polarization induced by the carrier in the channel of organic field-effect transistors (OTFTs). 
 
 \section{Experimental}
 \label{sec:experimental}
 \subsection{Fabrication procedure}
 For this work, top-contact thin ($5$ nm, $10$ nm and $100$ nm thick active layer) pentacene transistors were fabricated by standard high-vacuum deposition techniques as we reported previously. \cite{pratontep04,daraktchiev05} Both thermal SiO$_2$ ($200$ nm thick) on heavily-boron-doped silicon wafer and single-crystal Al$_2$O$_3$ (sapphire) were used as the device gate-dielectric.\\
 
 \begin{figure}[!h]
  \centering
  \psfrag{dt}[][]{\colorbox{white}{$\nabla T$}}
  \psfrag{t1}[][]{$T_1$}
  \psfrag{t2}[][]{$T_2$}
  \psfrag{t3}[][]{$T_3$}
  \psfrag{v}[][]{$\Delta V$}
  \psfrag{t}[][]{$\Delta T$}
  \psfrag{cryostat}[][]{\colorbox{white}{Cryostat}}
  \psfrag{sample}[][]{Sample}
  \psfrag{l}{$L=600\,\mathrm{\mu m}$}
  \psfrag{vacuum-chamber}[][]{Vacuum Chamber}
  \includegraphics[scale=1.3]{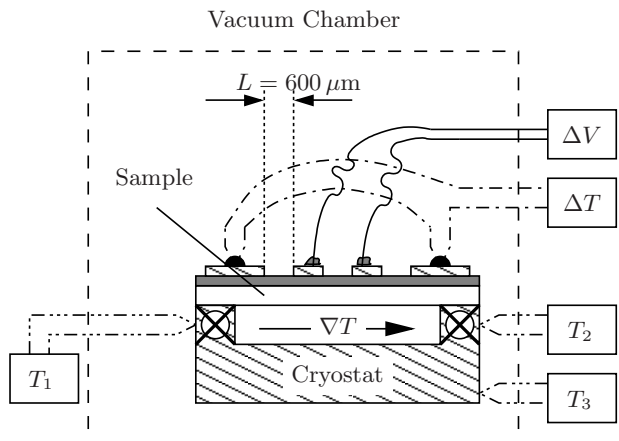}
  \caption{Schematic diagram of thermoelectric measurement apparatus. The sample was fixed by a Apezion\textregistered {} high vacuum grease to the copper heating elements to ensure the thermal contact. The thermocouples were cemented with Stycast\textregistered {} on the outer gold pads. The electromotive force $\Delta V$ was measured on the inner gold pads. The gold wires were glued with standard colloidal silver on the gold contacts. Thermocouples $T_1$, $T_2$ and $T_3$ were used to measure the temperature of the heating element and of the copper  mass (cryostat). The cryostat was electrically insulated and normally grounded ensuring $V_G = 0\, \mathrm{V}$.}
  \label{fig:schema}
 \end{figure}
 
 We also explored possible interfacial effects associated with the surface treatment of the oxide layer under various plasma exposure conditions and by vacuum deposition of a self-assembled monolayer (SAM) \cite{nuesch03} prior to pentacene growth. Plasma treatment with argon was done at a radio frequency of $10$ W for $5$ minutes at a constant gas pressure of $0.1$ mbar. The oxygen plasma treatment was done under the same conditions except the exposure time was set to $0.25$, $2$, $3$ and $10$ minutes.\\ 
 
 To form large pentacene grains within the conducting channels of the transistors, a substrate temperature of $328$ K and a deposition rate of $1.2$ nm/min were maintained during growth. \cite{daraktchiev05,pratontep04}
 The top source and drain gold contacts were evaporated through a shadow mask at the temperature of liquid nitrogen. The channels were fabricated with a width of $6 \, \mathrm{mm}$ and a length of either $50$ or $100$ $\mathrm{\mu m}$. The four-probe contact geometry consisted of four $6 \, \mathrm{mm}$-wide gold strips spaced $600$ $\mathrm{\mu m}$ apart. The morphology of the evaporated films was studied by atomic force microscopy (NT-MDT Solver Pro in tapping mode). 

 \subsection{Characterization}

 Dark current versus voltage ($I-V$) characteristics were obtained under ambient conditions using a Keithley 4200 semiconductor characterization system. The thermoelectric, four-probe and activated mobility measurements were carried-out under high vacuum of $1 \times 10^{-6} \, \mathrm{mbar}$ with a home made vacuum chamber.\\
 
 Thermoelectric power was measured by fixing the sample to two copper heating blocks (see Figure \ref{fig:schema}). The temperature was measured with a differential NiCr/CuNi thermocouple adhered to the gold contacts with a thermally stable resin. A temperature gradient of $5$ K along the sample was imposed for the duration of $1$ minute. The Seebeck coefficient was derived from the electromotive force $\Delta V$ divided by the applied temperature gradient $\Delta T$. To ensure that there were no artifacts due to the experimental setup, the direction of the temperature gradient was reversed and the measurement was repeated.

\section{Results}
\label{sec:results}
 \subsection{Field-effect mobility and activation energy}
\begin{figure}[!h]
  \centering
  \includegraphics{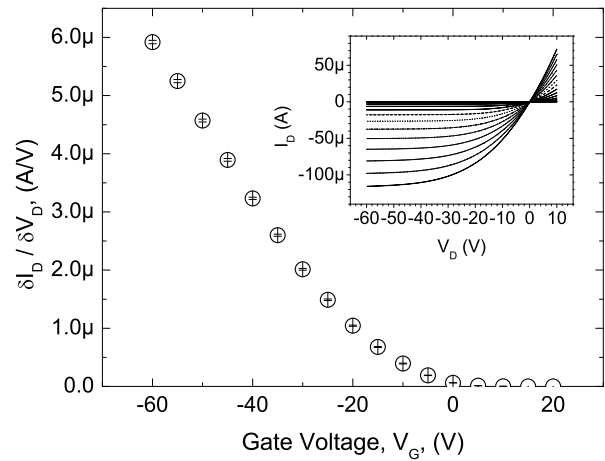}
  \caption{Transfer characteristics of transistor $9$ generated from the linear regime of the $I-V$ data in the inset. The field-effect mobility is $\mu=0.11  \, \mathrm{cm^2/V.s}$.}
  \label{fig:trans}
 \end{figure}

 The electrical characteristics of an OFET provide insight into the transport mechanisms via the field-effect mobility $\mu$, the activation energy of the mobility $\mathscr{E}_\mu$, and the activation energy of the thin-film conductivity $\mathscr{E}_a$.\\
 The $I-V$ characteristics of our pentacene thin-film transistors yielded field-effect mobilities in the range of $0.1$ to $0.4 \, \mathrm{cm^2/Vs}$ as shown in Table \ref{tab:mobility}. These mobility are deduced from the linear region of the $I-V$ characteristics. The activation energies are calculated from the slope of the associated Arrhenius plot. A typical result is given and analyzed in Figure \ref{fig:trans}.\\
 
 \begin{table}[!h]
  \begin{tabular}{|c|c|c|c|c|c|}
   \hline
    & Surface & 5acene & $\mu$ & $\mathscr{E}_a$ & $\mathscr{E}_{\mu}$\\
    & treatment & (nm) & $(\mathrm{cm^2/Vs})$ & (meV) & (meV)\\
   \hline
    $1$ & O$_2$ $180$ sec & $10$ & $0.22$ & $104$ & $95$ \\
    $2$ & Ar $300$ sec & $10$ & $0.32$ & $ 80$ & $32$ \\
    $3$ & O$_2$ $300$ sec & $5$ & $0.33$ & $-$ & $58$ \\
    $4$ & O$_2$ $300$ sec & $100$ & $0.2$ & $84$ & $-$ \\
    \multirow{2}{*}{$9$} & O$_2$ $120$ sec & \multirow{2}{*}{$10$} & \multirow{2}{*}{$0.11$} & \multirow{2}{*}{$161$} & \multirow{2}{*}{$107$}\\
    & COOH-3acene & & & &\\
   \hline
  \end{tabular}
  \caption{The field-effect mobility $\mu$ and the activation energies of the carrier conductivity $\mathscr{E}_a$ and the mobility $\mathscr{E}_\mu$ of our pentacene thin-film transistors.}
  \label{tab:mobility}
 \end{table}
 
\begin{figure}[!h]
  \centering
  \includegraphics{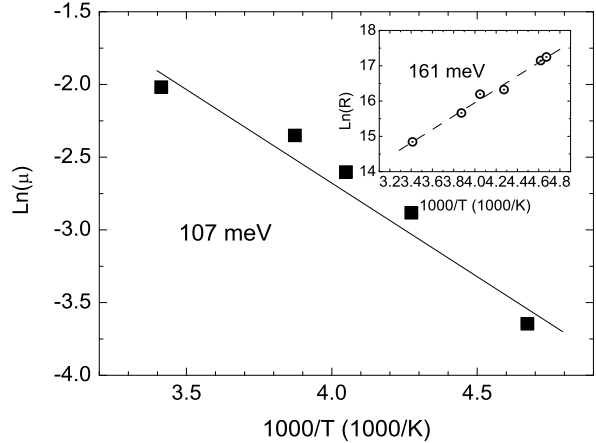}
  \caption{On transistor $9$, the field effect mobility is activated with the energy $\mathscr{E}_\mu = 107 \, \mathrm{m eV}$. The inset presents the four probe measurement of the temperature dependence of the channel conductivity. It is activated with $\mathscr{E}_a = 161 \, \mathrm{m eV}$ . The differences between the Fermi level $\mathscr{E}_F$ and the transport level $\mathscr{E}_V$ is then $\mathscr{E}_F - \mathscr{E}_V = 56 \, \mathrm{m eV}$.}
  \label{fig:activation_oxygen}
 \end{figure}

 The thin-film conductivities $\sigma$ were measured on the same substrate (from the same batch) by means of temperature-dependent four-probe technique. The activation energy is then derived from the slop of an Arrhenius plot as shown in Figure \ref{fig:activation_oxygen}. Both, the field-effect mobility and the channel conductivity exhibit typical temperature-activated behavior. This suggest that charge transport in organic thin-film transistors occurs via hopping between localized states. The significance of our channel conductivity results is discussed in more detail in Section \ref{sec:discussion}.
 
 \subsection{The thermoelectric power}
 For the thermoelectric experiment, the quantity of interest is the Seebeck coefficient ($\alpha$) which measures the entropy transferred to the carrier in the channel per unit charge. It is defined as $(\Delta V / \Delta T)$, where $\Delta V$ is the voltage induced by an applied temperature gradient $\Delta T$ across the sample. We can neglect the contribution of the contacts to the thermoelectric power because the measured values are two orders of magnitude higher than that of gold. Since we measured the electromotive force $\Delta V$ in open circuit, no electrical current was able to pass through the thin-film. Therefore the measurement is independent of the geometrical shape of the contacts and of the contact resistance and grain boundaries do not have a significant influence on the results.\\
 
\begin{figure}[!h]
  \centering
  \includegraphics{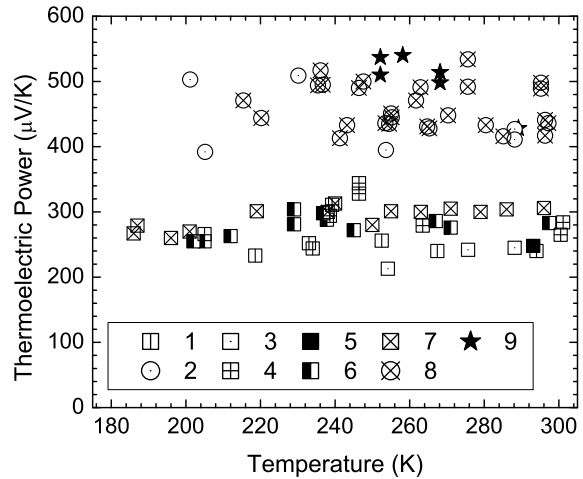}
  \caption{Seebeck coefficients measured on pentacene films deposited on the different substrates described in Table \ref{tab:summary}.}
  \label{fig:thermopower_total}
 \end{figure}

 The Seebeck coefficients that we obtained from our devices are presented in Figure \ref{fig:thermopower_total}.  We find surprisingly high values of $\alpha$ ranging from $240$ to $500$ $\mathrm{\mu V/K}$ at room temperature that depend on the nature of the device substrate.\\
 
 Although the results in Figure \ref{fig:thermopower_total} show a strong dependence of the thermoelectric power on the treatment of the oxide surface, we do not observe a dependence on the contact resistance or the morphology of the pentacene layer.  Table \ref{tab:thickness} gives the measured Seebeck coefficients of pentacene on oxygen plasma activated Si/SiO$_2$ substrates at room temperature for pentacene layer thicknesses of $5$, $10$, and $100$ nm.  It has been shown that the contact resistance of top-contact pentacene field-effect transistors increases with increasing active layer thickness. \cite{daraktchiev05} Since we observe little or no dependence of the thermoelectric power on the device thickness, we conclude that our measurements are independent of the contact resistance.
 
 \begin{table}[!h]
  \centering
  \begin{tabular}{| c | c | c | c |}
   \hline
   Sample & $3$ & $1$ & $4$ \\
   \hline 
   Thickness (nm) & $5$  & $10$   & $100$ \\
   Seebeck coefficient $\mathrm{(\mu V/K)}$  & $245$ & $240$ & $249$\\
   \hline
  \end{tabular}
   \caption{Thermopower measurements at room temperature for different sample thicknesses.}
  \label{tab:thickness}
 \end{table}

 \begin{figure}[!h]
  \centering
  \includegraphics{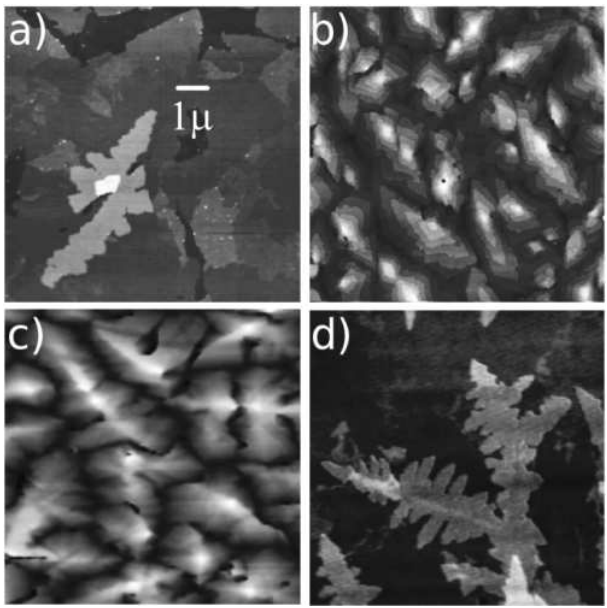}
  \caption{The topography of the pentacene films with various film thicknesses. Pentacene films on oxygen plasma activated Si/SiO$_2$ substrates are shown with thicknesses of: a) $5$ nm, b) $10$ nm, and c) $100$ nm. In panel d, the gate dielectric was modified by a SAM (COOH-Anthracene) prior to pentacene evaporation. The pentacene layer thickness in panel d is $10$ nm. All scans have dimensions of $10 \times 10 \, \mathrm{\mu m^2}$.}
  \label{fig:afm}
 \end{figure}

 The thermoelectric power is also independent of the morphology of the pentacene surface prior to gold contact deposition.  Figure \ref{fig:afm} shows AFM images of the pentacene surface in the device channel for various film thicknesses.  The pentacene films shown in panels a, b, and c were evaporated on top of Si/SiO$_2$ substrates treated with oxygen plasma and are $5$, $10$, and $100$ nm thick, respectively.  Panel d shows a $10$ nm-thick pentacene layer on top of a Si/SiO$_2$/Al$_2$O$_3$ substrate. The substrate was treated with oxygen plasma and then modified by a COOH-Anthracene SAM. As the film thickness increases, growth is less homogeneous and the morphology of the pentacene surface is more three-dimensional. However, given the lack of thickness dependence we observe in the Seebeck coefficient (Table \ref{tab:thickness}), these changes in morphology do not affect the thermoelectric power. This conclusion and the lack of observed contact resistance are consistent with previous results that show that charge transport occurs in the first few molecular monolayers adjacent to the gate dielectric. \cite{houili06,daraktchiev05}
 
\begin{figure}[!h]
  \centering
  \includegraphics{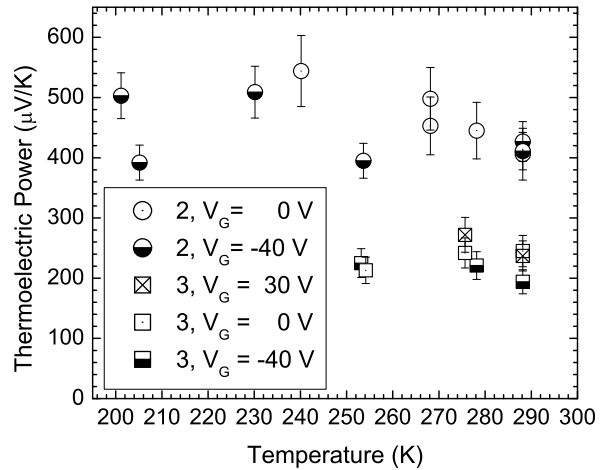}
  \caption{The Seebeck coefficient as a function of temperature $T$ and applied gate voltage $V_G$. Transistor $2$ was treated with argon plasma and transistor $3$ was treated with oxygen plasma.}
  \label{fig:gate_bias}
 \end{figure}

 It is also important to point out that the thermoelectric power does not depend on the gate bias strength. We measured the thermoelectric power of an argon plasma treated (transistor $2$) and an oxygen plasma treated device (transistor $3$) while applying a gate potential. The obtained Seebeck coefficients reported in Table \ref{tab:gate_bias} and Figure \ref{fig:gate_bias} do not exhibit a significant dependence on the applied gate potential. This confirms that the Seebeck coefficient in our devices is essentially a single carrier property. \cite{fritzsche1971} This is true in general except in the case of highly correlated electron systems, where it may depend slightly on the carrier density. \cite{chaikin1976}
  
 \begin{table}[!h]
  \centering
  \begin{tabular}{| c | c | c | c |}
   \hline
   & \multicolumn{3}{| c |}{$\alpha$ $\mathrm{(\mu V/K)}$ at room temperature}\\
   \hline
   Sample & $ V_G = -40 \, \mathrm{V}$ & $V_G = 0 \, \mathrm{V}$ & $V_G = +30 \, \mathrm{V}$\\
   \hline
   $2$ & - & $419$ & $406$ \\
   $3$ & $221$ & $245$ & $237$ \\
   \hline
  \end{tabular}
  \caption{Dependence of the Seebeck coefficient $\alpha\mathrm{(\mu V/K)}$ on the the applied gate bias $V_G$.}
  \label{tab:gate_bias}
 \end{table}
 
 \subsection{Discussion}
 \label{sec:discussion}
 We shall now examine the factors that contribute to such high values of $\alpha$.\\
 
 Following David Emin, \cite{emin1999} we recall that the main contributions to the Seebeck coefficient may be written as the sum of a formation contribution $\alpha_f$ and a migration contribution $\alpha_m$.

\begin{equation}
 \alpha = \alpha_f + \alpha_m
 \label{eqn:sum_thermopwr}
\end{equation}

 Here, $q \alpha_f$ is the change in the system's entropy induced by adding a charge carrier and is independent of the mechanism by which a carrier is transported through a material. On the other hand, $\alpha_m$ comes from the entropy change associated with the net energy transferred in moving a carrier. This contribution, which depends on the charge transport mechanism, was calculated by Emin \cite{emin1975} for the case of hopping transport in disordered systems. In our case, $\alpha_m$ is small when compared to $\alpha_f$. This is because of the nearly equal probability for upwards hops (involving energy gain) and downwards hops (involving energy loss) along the carrier hopping trajectory, leading to very small values of $\alpha_m$ of the order of a few ten of $\mathrm{\mu V/K}$. \cite{emin1975,overhof1975} If we assume further that spin contributions can also be neglected because carrier formation in organic semiconductors does not imply spin reorganizations, we are finally left with two contributions that we can write as,

 \begin{equation}
  \alpha \simeq \alpha_f = \frac{1}{q} \left (\Delta S_\text{mixing} + \Delta S_\text{vibration} \right )
  \label{eqn:Seebeck}
 \end{equation}

 The configuration entropy, $\Delta S_\text{mixing}$ reflects the changes in the distribution of the carriers between the electronic states. The vibrational entropy, $\Delta S_\text{vibration}$, emanates from the modification of the vibrational modes due to lattice softening or hardening induced by the presence of a charge carrier.\\

\begin{figure}[!h]
 \centering
 \includegraphics{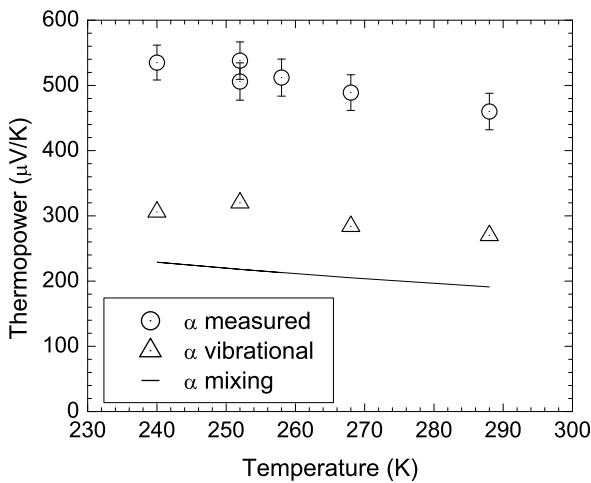}
 \caption{The temperature dependence of the mixing and vibrational contributions to the Seebeck coefficient for sample $9$.} 
 \label{fig:thermo_cooh}
\end{figure}

\begin{figure}[!h]
 \centering
 \includegraphics{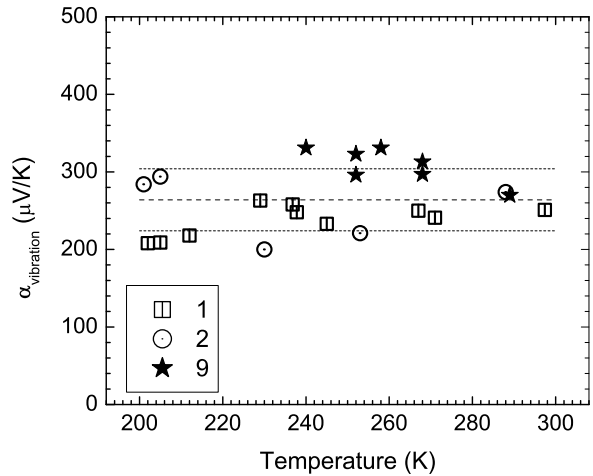}
 \caption{The contribution to the Seebeck coefficient due to interaction of the charge carrier with its surrounding phonons is about $265 \pm 40 \, \mathrm{\mu V/K}$ and \emph{independent} of the nature and the kind of surface treatment.}
 \label{fig:phonon_contribution}
 \end{figure}

 For non-degenerate hole systems, $\Delta S_\text{mixing}$ is well-known \cite{heikes1961,fritzsche1971} and can be approximated by,

 \begin{equation}
  \Delta S_\text{mixing} = \left \langle \frac{\mathscr{E}_F - \mathscr{E}}{T} \right \rangle \simeq \frac{\mathscr{E}_F -\mathscr{E}_V}{T}
  \label{eqn:mixing_entropy}
 \end{equation}
 
\noindent where $\mathscr{E}_F$ represents the position of the Fermi level, $\mathscr{E}$ is the hole energy and $\mathscr{E}_V$ is the transport energy. The bracket corresponds to thermal average over all the electronic states which contribute to conduction. We can estimate the value of $\Delta S_\text{mixing}$ by extracting $\Delta \mathscr{E} = \mathscr{E}_F - \mathscr{E}_V$ from four-probe measurement of the temperature dependence of the channel conductivity. A typical result is shown in Figure \ref{fig:activation_oxygen}. The quasi-two dimensional channel conductivity portrays an activated behavior arising from the thermal excitation of the charge carrier from the transport energy to the Fermi level and from carrier hopping between localized states. We modelled this previously \cite{daraktchiev05} as,

 \begin{align}
  \sigma & = |e| p_0 \exp{\left\{ \frac{-\Delta \mathscr{E}}{k_\text{B}T} \right\} } \mu_0 \exp{\left\{ \frac{-\mathscr{E}_{\mu}}{k_\text{B}T} \right\} }\notag \\
   & = \left | e \right | p_0 \mu_0 \exp{\left \{ \frac{- \mathscr{E}_a}{k_\text{B} T}\right \} } 
  \label{eqn:activation_energy} 
 \end{align}
              
 \noindent where  $ \mathscr{E}_a = \Delta \mathscr{E} + \mathscr{E}_\mu$  and  $\mathscr{E}_\mu$ is the activation energy associated with the field-dependent mobility.\\
 From this analysis, we determine the vibrational contribution to the measured Seebeck coefficients shown in Figure \ref{fig:thermo_cooh} as $\alpha_\text{vibrational} = \Delta \alpha + \alpha_\text{mixing}$. Table \ref{tab:summary} summarizes the results of our analysis. As displayed in Figure \ref{fig:phonon_contribution}, we find an unusually large vibrational contribution to $\alpha$ of about $265 \pm 40 \, \mathrm{\mu V/K}$ that is independent of the type of gate-dielectric and the different surface treatment performed on the substrates.

 \begin{table}[!h] 
  \centering
  \begin{tabular}{|c|c|c|c|c|}
   \hline
   & Substrate & Surface & 5acene & $\alpha$ \\ 
   & & treatment & (nm) &  ($\mathrm{\mu}$V/K)\\ 
   \hline
   $1$ & Si/SiO$_2$ & O$_2$ $180$ sec & $10$ & $240$\\
   $2$ & Si/SiO$_2$ & Ar $300$ sec & $10$ & $419$\\
   $3$ & Si/SiO$_2$  & O$_2$ $300$ sec & $5$ & $245$\\
   $4$ & Si/SiO$_2$ & O$_2$ $300$ sec & $100$ & $249$\\
   $5$ & Al$_2$O$_3$ & O$_2$ $15$ sec & $10$ & $284$\\
   $6$ & Al$_2$O$_3$ & O$_2$ $180$ sec & $10$ & $283$\\
   $7$ & Al$_2$O$_3$ & O$_2$ $600$ sec & $10$ & $306$\\
   $8$ & Al$_2$O$_3$ & Ar $300$ sec & $10$ & $429$\\
   \multirow{2}{*}{$9$} & \multirow{2}{*}{Si/SiO$_2$/Al$_2$O$_3$} & O$_2$ $120$ sec &  \multirow{2}{*}{$10$} & \multirow{2}{*}{$460$}\\
   & & COOH-3acene & & \\
   \hline 
   \hline
   & $\mathscr{E}_{\mu}$ &  $\mathscr{E}_F - \mathscr{E}_V$\footnotemark &  $\Delta  \alpha_\text{mixing}$ & Difference\\
   & (meV) & (meV) & ($\mathrm{\mu}$V/K) & ($\mathrm{\mu}$V/K)\\
   \hline
   $1$ & $95$ & $9$ & $32$ & $251$\\
   $2$ & $32$ & $38$ & $172$ & $274$\\
   $9$ & $107$ & $56$ & $191$ & $270$\\
   \hline
  \end{tabular} 
  \caption{Summary of the achieved thermoelectric power upon different substrate and surface treatments.} 
  \label{tab:summary}
  \footnotetext{at $300$ K} 
 \end{table}

\section{Model}
 We now present a model calculation that accounts for this very large intrinsic value of  $\alpha_\text{vibrational}$ in OTFTs.\\
 When phonons with energies lower than $k_\text{B}$ are involved, the vibrational  entropy (which in fact corresponds to the specific heat) follows the law of Dulong and Petit. It leads to a temperature independent contribution \cite{mccready01} given by,
 
 \begin{equation}
  \Delta S_\text{vibration} = k_\text{B} \left [ \sum_j N_j \left | \frac{\Delta  \nu}{\nu} \right | _j \right]
  \label{eqn:vibrational_enthropy}
 \end{equation} 

 \noindent where $N_j$ represents the number of softened (or hardened) modes, and $\nu_j$ their frequency.\\

 Equation \eqref{eqn:vibrational_enthropy} relates the thermopower vibrational contribution to the local changes in the phonon frequencies induced by a charge. This contribution is exceptionally large for the highly polarizable organic molecular semiconductors \cite{picon04} as the nature of the chemical bonding is changed by the presence of a charge carrier from pure Van-der-Waals to charge-dipole interaction. More precisely, if the carrier is localized on a molecule, the potential energy in which the adjacent molecules move, is the sum of the potential energy arising from direct interactions between neutral molecules and the carrier electronic energy, itself a function of the molecule positions $\vec{r_i}$. It is given by,  

 \begin{displaymath}
  \mathscr{E}_p(\vec{r_i}) = -\frac{1}{2} \sum_i q \frac{\vec{r_i} \cdot \vec{d_i}}{4 \pi \varepsilon_0 r_i^3}
 \end{displaymath}

 \noindent where $\vec{d_i}$ is the dipole induced on the $i^\text{th}$ molecule by the carrier at the origin. \cite{picon04} Their contribution to the polarization energy $\mathscr{E}_p(\vec{r_i})$ decreases as the fourth power of their distance from the carrier. Then, we evaluate the local change in the phonon frequencies in the spirit of a pair-potential model. We start with a Lennard-Jones potential and a Van-der-Waals attractive potential for a pair of neutral molecules and we add the contribution of the carrier electronic energy, \cite{picon04}

 \begin{eqnarray}
  \mathscr{V}(r) = \frac{-M \omega^2 r_0^2}{72} \left[ 2 \left( \frac{r_0}{r} \right)^6 - \left(\frac{r_0}{r} \right)^{12} \right] \nonumber\\
  - \frac{\gamma}{2} \frac{q^2}{(4 \pi \varepsilon_0)^2 r_0^4}  \left( \frac{r_0}{r} \right)^4
  \label{eqn:potential}
 \end{eqnarray}

 \noindent where $M$ is the reduced mass of the molecule, $r_0$ the equilibrium distance between neutral molecules, $\gamma$ the electronic polarizability of the molecule, $q$ is the charge of the carrier. The shift in the carrier equilibrium position from $r=r_0$ to $r = r_0^*$ is obtained from the solution of $(\partial \mathscr{V}/\partial r)_{r=r_0^*} = 0$. The result is,
 
 \begin{align}
  \frac{r_0}{r_0^*} & = \left( 1 + \frac{12 \gamma q^2}{(4 \pi \varepsilon_0 )^2 r_0^4} \frac{1}{M \omega^2 r_0^2} \right )^{1/6}
  \label{eqn:sol_potential} 
 \end{align}
 
 The stiffness parameter is modified from $k = M \omega^2$ to $k^* = k + \delta k = (\partial^2 \mathscr{V}/\partial r^2)_{r=r_0^*}$ and we find
 
 \begin{displaymath}
  k^* = M \omega^2 \left[ 1 + \frac{32 \gamma q^2}{(4 \pi \varepsilon_0)^2 r_0^4}  \frac{1}{M \omega^2 r_0^2} \right]
 \end{displaymath}

 For a pair of pentacene molecules, these parameters are estimated as follows: $\gamma / 4 \pi \varepsilon_0 = 40 \text{ \AA}^3$, $r_05 \text{ \AA}$, $M = 23 \times 10^{-26} \text{ kg}$ and $\omega = 2\pi \nu \sim 10^{13} \text{ s}^{-1}$, so that $\delta \nu / \nu =(1/2) (\delta k/k)  = 0.32$. Since the frequency hardening is proportional to $r_0^{-6}$, the strongest contribution to the thermopower comes from the four next-nearest neighbors of the ionized molecule in the pentacene crystal. In our case, by considering two optical modes in the plane, the estimated thermopower is
 
 \begin{equation}
  \alpha_\text{vibration} \sim 8 \frac{k_\text{B}}{q} \frac{\Delta \nu}{\nu} \simeq 220 \, \mathrm{\mu V/K}
  \label{eqn:prediction_thermopower}
 \end{equation}

 \noindent This calculated value agrees remarkably well with the measured values reported in Figure \ref{fig:phonon_contribution}.\\

\section{conclusion}
 The above results strongly support the conclusion that an electronic polaron is formed in the quasi-two dimensional channel of organic field-effect transistors. In the pentacene thin-films presented here, the charge carriers are localized due to the presence of interfacial disorder. Thus, the electronic polaron is quasi-static.\\
 
 In high mobility transistors based on pentacene single crystals \cite{morpurgo04,podzorov01} the electronic polaron is delocalized as Bloch waves. \cite{picon04} As further corroboration of our results, we predict the vibrational entropy to be weaker in single-crystals than in thin-films. Understanding the dynamics of the electronic polaron motion can lead to quantitative assessment of the observed charge transport in organic molecular semiconductors and the optimization of the performance of optoelectronic devices based on them. We have recently proposed a new theory to predict the mobility of this electronic polaron. \cite{picon06}

\begin{acknowledgements}
 We wish to thank our colleagues, M. Castellani for scientific support, M. Longchamp and C. Amendola for technical help, S. J. Konezny and D.B. Romero for scientific discussions and the Swiss National Science foundation for financial support (project number 200020-113254).
\end{acknowledgements}

\end{document}